\documentclass[prl,nofootinbib,
noshowkeys,twocolumn,superscriptaddress]{revtex4}
\usepackage{graphicx}

\newcommand{\be}{\begin{eqnarray}}
\newcommand{\ee}{\end{eqnarray}}

\newcommand{\ra}{\rangle}

\begin{document}
\author{Pietro Faccioli}
\email{faccioli@ect.it}
\affiliation{ E.C.T.*, Strada delle Tabarelle 286, 
I-38050 Villazzano (Trento)}
\affiliation{ INFN, Sezione di Trento, Trento, Italy.}
\author{Thomas A. DeGrand}
\email{thomas.degrand@colorado.edu}
\affiliation{ Department of Physics,  University
 of Colorado, Boulder, CO 80309, USA\cite{ref:present}}
\affiliation{ Max-Planck-Institut f\"ur Physik (Werner-Heisenberg Institut),
F\"ohringer Ring 6, 80805 M\"uncnen, Germany}
\title{Evidence for  Instanton-Induced Dynamics, from Lattice QCD}
\vspace {2cm}
\begin{abstract}

We perform a study of the non-perturbative dynamics of the
light-quark sector of QCD, based on
some recent results of lattice simulations with chiral fermions.
We analyze some correlators that are designed to probe 
the Dirac structure of the quark-quark interaction at different scales.
We show that, in the non-perturbative regime, such an interaction
contains very large scalar and pseudo-scalar components.
We observe {\it quantitative} agreement between lattice QCD results
and the predictions of  the Instanton Liquid Model (ILM). 
Moreover, we study how the
the quark-quark interaction is modified, when quark loops are suppressed.  
We observe a dramatic effect related to the loss of unitarity, which 
is naturally explained in the instanton picture. 
Such an effect cannot be explained in a Dyson-Schwinger
Equations (DSE) approach, if one assumes a vector quark-gluon coupling. 
Therefore, from the present study it emerges that DSE models 
with such an ansatz for the vertex function are not consistent with QCD.
 
\end{abstract}
\maketitle
%

The physics of the light-quark sector of QCD is strongly influenced 
by the non-perturbative structure of the vacuum, and in particular by
the spontaneous breaking of chiral symmetry (SCSB).
Hence, identifying the microscopic mechanism responsible for such a phenomenon
represents a fundamental step toward the comprehension of the strong
interaction. 
Unfortunately, this dynamics 
resides in the non-perturbative sector
of QCD and has not been completely understood from first principles.

The typical energy scale of phenomena related to the breaking of chiral 
symmetry is  $4\pi\,f_\pi\simeq\textrm{1.2\,GeV}$ , 
is considerably larger than
the confinement scale, $\Lambda_{QCD}\sim 1\textrm{fm}^{-1}\sim
0.2~\textrm{GeV}$. 
From a theoretical perspective, such a separation of scales is crucial, 
because it justifies attempting  
model descriptions of the non-perturbative physics of SCSB, 
without needing to simultaneously 
take into account the dynamics of quark confinement.
The common feature in all these semi-phenomenological approaches is a  
strong attraction in the flavor-singlet $O^+$ channel, leading
to a quark condensate.
On the other hand,
 some of the models which have been proposed in the literature
rely on drastically different microscopic
descriptions of the non-perturbative quark-quark interaction.

Historically, the first attempt to explain the breaking of chiral symmetry
predates QCD and was developed in the
Nambu-Jona Lasinio  model \cite{NJL}, where a chirally-symmetric effective
Lagrangian, characterized by 
a scalar and pseudo-scalar four-fermion interaction, 
was postulated. Later on, the same structure was recovered in 
the context of the ILM\cite{shuryakrev}. 
The latter approach has the advantage of
being formulated in terms of quark and gluon
degrees of freedom and to be motivated from QCD through a 
semi-classical argument. Moreover, it explains in a very
natural way the structure of the 
spectrum of lowest-lying eigenvalues of the Dirac operator.
\begin{figure}
\includegraphics[scale=0.5,clip=]{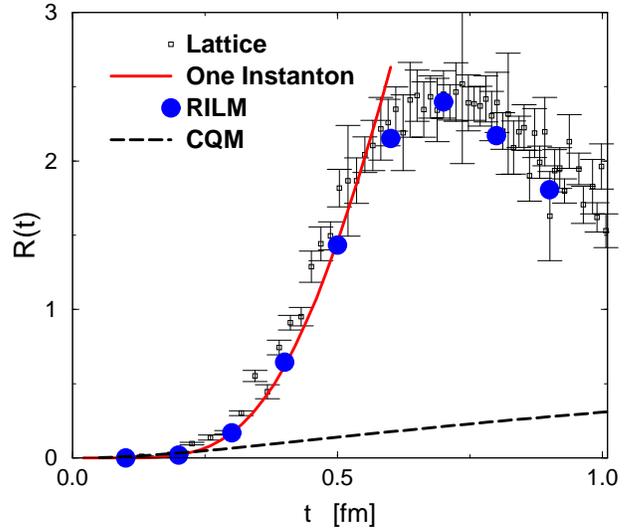}
\caption{The chirality-flip ratio, $R^{NS}(\tau)$, in lattice and
in two phenomenological models.
Squares are lattice points of \cite{degrand1}. 
Stars are RILM points
obtained numerically from an ensemble of 100 instantons of 1/3 fm size in a 
$5.3\,\times\,2.65^3\,\textrm{fm}^4$ box.
The solid line is the contribution of a single-instanton, 
calculated analytically in \cite{chimix}.
The dashed curve was obtained from  two free ``constituent''
quarks with a mass of
400 MeV. Such a curve {\it qualitatively} resembles the
prediction of a model in which chiral symmetry is broken through a vector
coupling (like in present DSE approaches).}
\label{scalar}
\end{figure}

An alternate model description of the non-perturbative sector of QCD which
encodes the physics of SCSB has been developed in the context 
of Dyson-Schwinger Equations (DSE). 
In such an approach, one parametrizes the 
low-energy behavior of QCD 
through an ansatz of the infrared structure of the quark-gluon vertex
and of the propagators \cite{DSE1}.
DSE are then solved numerically, in a given truncation scheme.

Although both DSE on the one hand, and the ILM on the
other hand, give comparable phenomenology in the light
hadron sector, they rely on 
drastically different microscopic pictures of the non-perturbative 
interaction, at the 1 GeV scale. 
Most applications of the DSE developed so far assume a simple vector 
ansatz\footnote{For an example of a DSE model with a more general ansatz 
for the quark-gluon vertex, see Fischer and Alkofer\cite{alkhofer}.}   
for the quark-gluon vertex function, $\Gamma_\mu \propto \gamma_\mu$.
In other words, the non-perturbative dynamics is  
mediated by the exchange of {\it one} (albeit non-perturbative) 
gluon at the time.

On the other hand, in the instanton picture, 
the non-perturbative dynamics is dominated by the 't Hooft interaction. 
Through standard bosonization of the 't Hooft vertex, such an 
instanton-induced interaction can be thought as being mediated
by  fields carrying the quantum numbers of  scalar and pseudo-scalar bosons.

The goal of this Letter is to identify 
which one of these two alternate microscopic pictures is closer to QCD.
We shall present 
evidence for the existence of a large scalar and pseudo-scalar component
of the non-perturbative quark-quark low-energy interaction.
This evidence is built using some recent results from
lattice simulations with chiral fermions\cite{degrand1}.
On the one hand, these results agree on a {\it quantitative}
level with the predictions of the ILM.
On the other hand,  they rule-out any picture in which the non-perturbative 
quark-quark interaction is assumed to have a vector structure,
like in present DSE models.

Our analysis is based on the study of
the flavor Non-Singlet (NS) chirality-flip ratio,  introduced
in \cite{chimix}:
\be
\label{RNS}
R^{NS}(\tau):=\frac{A^{NS}_{flip}(\tau)}{A^{NS}_{non-flip}(\tau)}=
\frac{\Pi_\pi(\tau)-\Pi_\delta(\tau)}
{\Pi_\pi(\tau)+\Pi_\delta(\tau)},
\ee
where $\Pi_\pi(\tau)$ and $\Pi_\delta(\tau)$ are pseudo-scalar and scalar NS
two-point correlators\footnote{Notice that the above correlators are defined in
coordinate representation, 
they are not zero-momentum projections.} 
related to the currents $J_\pi(\tau):=\bar{u}(\tau)
\,i\gamma_5\,d(\tau)$ 
and $J_\delta(\tau):=\bar{u}(\tau)\,d(\tau)$.
If the propagation is chosen along the (Euclidean) 
time direction, $A^{NS}_{flip(non-flip)}(\tau)$ represents 
the probability amplitude for a 
$|q\,\bar{q}\ra$ pair with iso-spin 1 to be found after a 
time interval $\tau$ in a state
in which the chirality of the quark and anti-quark 
is interchanged (not interchanged).
Notice that the ratio $R^{NS}(\tau)$ 
must vanish as $\tau\to 0$ (no chirality flips),
 and must approach 1 as  
$\tau\to \infty$ (infinitely many chirality flips).

In \cite{chimix} it was shown that the correlator (\ref{RNS}) 
is a particularly useful theoretical tool for studying
the non-perturbative dynamics of the light quark sector of QCD.
In fact, $R^{NS}(\tau)$ receives no leading perturbative contribution and
probes directly the chirality-mixing interaction.
A spectral analysis of $R^{NS}(\tau)$
indicated that the such an interaction is
mediated by topological fields.
In particular, it was found that 
the rate of chirality flips in a quark-antiquark system
is proportional to the mass of the $\eta'$ meson.
Moreover, below we shall see that
$R^{NS}(\tau)$ is very sensitive 
to the Dirac structure of the non-perturbative quark-quark interaction.

The NS scalar and pseudo-scalar two-point functions
appearing in (\ref{RNS}) have been first calculated in the
quenched approximation by the MIT group \cite{lattice2pt} with 
Wilson fermions and more 
recently  by one of the authors,
using chiral (overlap) fermions \cite{degrand1}.
The curve for $R^{NS}(\tau)$ obtained 
from the result of latter calculation\footnote{We recall that, 
with overlap fermions,  the lattice to continuum renormalization 
factors of the pseudoscalar and scalar correlators are equal, 
and drop out in the ratio.}  
are the square points plotted in Fig.~\ref{scalar}.

As expected, the lattice data interpolate between 0 and 1. Notice that
the curve has a maximum
at about 0.7~fm, where the ratio is considerably larger than one.
This implies that, after few fractions of a fermi, 
the quarks are more likely to be found
in a configuration in which their chiralities is flipped, than to be
found in their initial 
configuration.
Below we shall see that the presence of such a maximum is a signature
of a chirality mixing component of the quark-quark effective 
interaction vertex.

We recall that these lattice 
results have been obtained in the quenched approximation. 
It is important to ask what differences should be expected in full QCD. 
Using general QCD inequalities \cite{inequalityQCD}, 
it is immediate to show that $R^{NS}(\tau)>1$ if and only if 
 $\Pi_\delta(\tau)<0$.
The negativity of such a two-point function represents a severe failure of the
quenched approximation which appears only at sufficiently small values
of the quark mass. (In the large mass limit, the quenched approximation
becomes exact.)
It is a reflection of the fact that,
in the quenched approximation,  the unitarity of the theory is lost.

In terms of chirality flipping amplitudes, 
we see that the $\Pi_\delta(\tau)\ge 0$ 
constraint implies that quarks must never be
more likely to  be found in the flipped chirality configuration 
than that in the original configuration, 
$A_{flip}(\tau)\le A_{non-flip}(\tau)$.
Hence,  we can conclude
that the fermionic determinant suppresses some
chirality flipping events, which are otherwise allowed in the
quenched approximation.
Indeed, the correlators appearing in (\ref{RNS}) have recently been evaluated
in full QCD, with Wilson fermions \cite{Martinez}. 
It was observed that
the condition $\Pi_\delta(\tau)>0$ (or, equivalently, $R^{NS}(\tau)<1$)
is restored in going from quenched to full QCD.
Such a dramatic qualitative difference between quenched and full QCD
calculations of (\ref{RNS}) can be used to test phenomenological descriptions 
 of the non-perturbative dynamics. 
 Indeed {\it any realistic model must reproduce a dramatic enhancement
of the chirality flipping amplitude, when quark loops are suppressed.}

\begin{figure}
\includegraphics[scale=0.5,clip=]{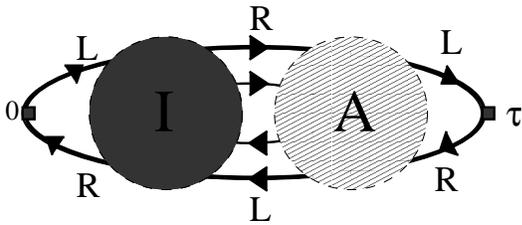}
\caption{Suppression of chirality flips due to the topological
screening induced by fermion-loops in the ILM ($N_f=3$).}
\label{quenching}
\end{figure}

Let us now discuss how  $R^{NS}(\tau)$ looks 
in the ILM and DSE models mentioned above.
In both approaches chiral symmetry is spontaneously 
broken, albeit through two very different microscopic mechanisms.
As a consequence, the 
quarks acquire dynamically an effective mass, which triggers 
chirality mixing.
Due to such a mass generation process, in both cases 
$R^{NS}(\tau)$ interpolates between 0 and 1.
However, one expects crucial {\it qualitative} differences
 between the prediction of the ILM and that of the DSE models\footnote{
More generally, we expect differences between the prediction of
any model with a vector structure of the vertex on the one hand, and
any model with a scalar and pseudo-scalar coupling on the other hand.}.
Let us first 
consider the result for  $R^{NS}(\tau)$ obtained 
in the Random Instanton Liquid Model\footnote{In the following discussion
one can regard the RILM as a prototype of a model with
a scalar and pseudo-scalar non-local interaction between quasi-zero modes.
The strength  of the coupling is tuned by the instanton density, while the 
delocalization of the quark-quark interaction is provided by the instanton 
size.}. In such a version of the ILM
quarks are assumed to propagate in a vacuum populated by an ensemble of
randomly distributed instanton and anti-instantons 
with a density of $\bar{n}=1\,\textrm{fm}^{-4}$ 
and size $\bar{\rho}=1/3\,\textrm{fm}$.
It can been shown that the RILM accounts for the 't Hooft interaction to
all orders, but neglects quark loops (quenched approximation).

The RILM prediction for $R^{NS}(\tau)$ is presented in Fig.~\ref{scalar}. 
The agreement with the lattice results is 
impressive\footnote{The RILM points presented in Fig.~\ref{scalar}
 have been obtained 
using a quark mass of about 30~MeV, that is the same as the same bare mass
used in one of the author's lattice simulations.}.
It is quite remarkable that not only does the RILM curve
displays a maximum in $R^{NS}(\tau)$, 
but also its position agrees quantitatively with the lattice results.

The presence of a maximum in $R^{NS}(\tau)$, 
and the subsequent fall-off towards 1 
have a very simple explanation in the RILM: 
if quarks propagate in the vacuum for a time comparable with
the typical distance between 
two neighbor instantons (i.e. two consecutive 't Hooft interactions), 
they have a large probability of crossing the field of the closest
instanton. 
If so happens, they must necessarily flip their chirality, due to the 
scalar and pseudoscalar structure of the 't Hooft vertex.
So, after some time, the quarks are most likely to be found in the
configuration in which their chirality is flipped.
On the other hand, 
if one waits for a time much longer than 1~fm, then the quarks will ``bump''
 into many other pseudo-particles,
experiencing several more chirality flips. Eventually, either chirality 
configurations will become equally probable 
and $R^{NS}(\tau)$ will approach 1.

The position of the maximum in $R^{NS}(\tau)$ carries
information about the interplay between one-body
and many-body effects generating chirality mixing.
It is interesting to compare the above numerical
RILM results with the single-instanton contribution (solid
line in Fig. \ref{scalar}), which
was derived analytically by one of the authors in \cite{chimix}.
From such a comparison, one can see that one-body effects,
i.e. chirality flips induced at the level of a {\it single} interaction, 
dominate the ratio up to Euclidean times of the order of 0.5-0.7 fm. 
The onset of many-body (many-instanton) effects is 
governed by the numerical value of the instanton density: 
the less dilute is the system, 
the earlier many-instanton effects become important. 
From the agreement between ILM and lattice data one may argue that
the phenomenological 
value $\bar{n}\simeq~1\,\textrm{fm}^{-4}$ is indeed realistic.

The prediction for $R^{NS}(\tau)$ would be
drastically different in the DSE models and in general
in all approaches based on a vector quark-gluon coupling.
In this case, the chirality mixing is only due to the dynamical
mass generation  (i.e. by a genuine many-body effect). 
In fact, unlike in the ILM,
a  {\it single} quark-antiquark interaction will not interchange the 
chirality of quarks, because the vector Dirac structure of the 
quark-gluon vertex is chirality conserving. 
As a result, even in the quenched approximation, 
quarks are {\it never} more likely to be found in the 
flipped chirality state than in the initial chirality state, i.e. 
$R^{NS}(\tau)<1$ for all $\tau$.
On a {\it qualitative} level, the DSE prediction for 
$R^{NS}(\tau)$ will be similar to that 
obtained considering the propagation
of a free but {\it massive} ``constituent'' quark and anti-quark pair in the
vacuum (dashed line in Fig.~\ref{scalar}).

From the above analysis we can conclude
that the presence of a maximum in the lattice results for
$R^{NS}(\tau)$ implies that 
the non-perturbative
quark-quark interaction contains a strong scalar and pseudo-scalar component.
In other words, the chirality is mixed already at the level
of a {\it single} quark-quark interaction, and not only through many-body 
effects, such as the mass generation induced by SCSB.
Therefore,  these lattice simulations
strongly support the ILM picture against the DSE models. 

\begin{figure}
\includegraphics[scale=0.5,clip=]{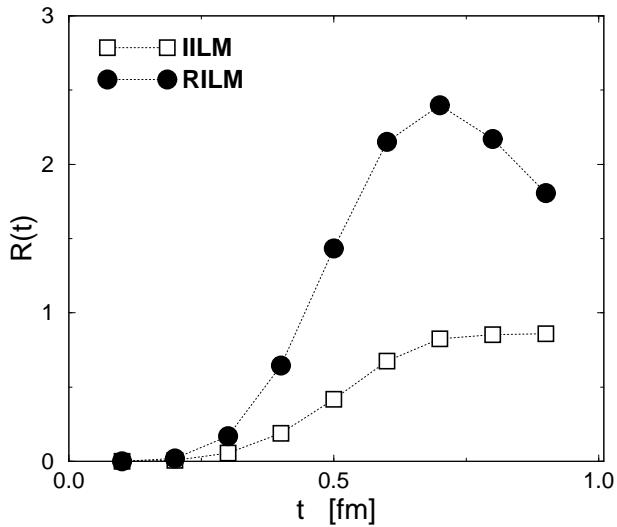}
\caption{Suppression of chirality flipping events, due to fermion-loop 
exchange in the ILM. Circles are RILM (quenched) results, squares are IILM 
(unquenched) results. In the unquenched model the unitarity requirement,
$R^{NS}(\tau)\le 1$ is restored.}
\label{unquenching}
\end{figure}
Additional evidence in this direction comes from comparing quenched 
and unquenched predictions.
In the DSE model, neglecting quark loops only affects the speed of the
running of the coupling, but does not generate additional chirality flips.
On the contrary, we already mentioned that
a dramatic qualitative difference between quenched and unquenched 
results is expected and observed on the lattice.
Such a difference is naturally explained in the ILM\footnote{P.F. acknowledges
a clarifying discussion with E.Shuryak on this point.}.  
If quark loops are allowed, then
instantons and anti-instantons can interact through
fermion exchange.
Such an interaction  generates an attraction between 
instantons and anti-instantons
leading to a screening of the topological charge \cite{toposcreening}
and providing a realization of the Leutwyler-Smilga  \cite{Leutwyler:1992yt}
relation (topological susceptibility
vanishing linearly with quark mass, at small quark mass).
As a result of such a screening, quarks crossing the field of an instanton 
are very likely to find, in the immediate vicinity, an
anti-instanton which restores their initial chirality 
configuration\footnote{From this discussion it follows that,
 in a correlated instanton vacuum,
the topological screening length coincides with the
typical time interval $\bar{\tau}$ between two consecutive 
chirality flipping interactions. 
Indeed, in \cite{toposcreening} such a screening length was estimated 
from numerical simulations in the ILM 
and was found to be $0.2-0.3$~fm, consistent
with the value $\bar{\tau}=1/m_{\eta'}\simeq~0.2$~fm, 
obtained by one of the authors in \cite{chimix}} 
(see Fig.~\ref{quenching}).

In Fig.~\ref{unquenching} we compare the chirality flipping ratio
 $R^{NS}(\tau)$ obtained from a quenched (RILM)
and unquenched (IILM\footnote{Interacting Instanton Liquid Model}) 
version of the ILM.
We observe that, with the inclusion of the fermionic determinant, 
the condition $R^{NS}(\tau)<1$ is restored. 
We  stress that, although such a restoration  
must necessarily take place in QCD, it represents a remarkable success 
of the ILM, which is not a unitary field theory.

In conclusion, we have presented a study of the microscopic 
structure of the non-perturbative 
interaction in QCD, based on the 
analysis of the results of some recent lattice simulations with chiral 
fermions.
We have used these data to test model descriptions of the microscopic
quark dynamics. 
We have found evidence for a large scalar and
pseudo-scalar component of the effective quark-quark vertex.
We have argued that this result rules out models in 
which the quarks couple non-perturbatively though a purely vector quark-gluon
vertex. 
On the contrary, we have observed impressive quantitative agreement 
between lattice and ILM results. In addition, the ILM
can also explain how quark loops cause a drastic
suppression of quark chirality flips.

\acknowledgements

It a pleasure to 
thank G. Ripka, T.Sch\"afer, E. Shuryak, and 
W. Weise for interesting
discussions and for critical reading of the manuscript. 
T. D. and P. F. would like to thank
respectively the Max-Planck-Institute in Munich and 
Physics Department of the Roma-Tre University for their 
hospitality while this paper was written. 
This work was partially supported by the U.~S. Department
of Energy, with grant DE-FG03-95ER40894.
{}
\end{document}